\documentclass[12pt]{nature}
\usepackage{amsmath,xcolor,graphicx,colortbl,multirow}
\usepackage{amssymb,soul,color}
\usepackage{booktabs}
\usepackage[left]{lineno}
\usepackage[version=3]{mhchem}
\usepackage{bm}
\usepackage{booktabs}
\usepackage{multirow}
\usepackage{float}
\usepackage{mwe}
\usepackage{subfig}
\usepackage{adjustbox}
\usepackage{graphicx}
\usepackage{dcolumn}
\usepackage{bm}

\usepackage[utf8]{inputenc}
\usepackage[T1]{fontenc}
\usepackage{mathptmx}
\usepackage{amsmath,xcolor,colortbl,multirow,chngcntr,float,xr}

\begin{document}

\title{Fine Structure of Excitons in Vacancy Ordered Halide Double Perovskites}
\author{Bruno Cucco,$^\dagger$ Claudine Katan,$^\dagger$ Jacky Even,$^\ddagger$ Mikael Kepenekian,$^{\dagger}$ George Volonakis$^{\dagger,*}$\vspace{0.2cm}}

\maketitle
\vspace{0.4cm}
\begin{affiliations}
 \item $^\dagger$~{Univ Rennes, ENSCR, INSA Rennes, CNRS, ISCR (Institut des Sciences Chimiques de Rennes), UMR 6226, France}
 \item $^\ddagger$~{Univ Rennes, INSA Rennes, CNRS, Institut FOTON - UMR 6082, Rennes, France}
 \item $^{*}$~{Corresponding author: yorgos.volonakis@univ-rennes1.fr}
\end{affiliations}

\begin{abstract}
Vacancy ordered halide double perovskites (VODP) have been widely explored throughout the past few years as promising lead-free alternatives for optoelectronic applications. Yet, the atomic-scale mechanisms that underlie their optical properties remain elusive. In this work, a throughout investigation of the excitonic properties of key members within the VODP family is presented. We employ state-of-art ab-initio calculations and unveil critical details regarding the role of electron-hole interactions in the electronic and optical properties of VODP. The materials family is sampled by picking prototypes based on the electronic configuration of the tetravalent metal at the center of the octahedron. Hence, groups with a valence comprised of s, p and d closed-shells are represented by the known materials Cs$_{2}$SnX$_{6}$, Cs$_{2}$TeX$_{6}$ and Cs$_{2}$ZrX$_{6}$ (with X=Br, I), respectively. The electronic structure is investigated within the G$_{0}$W$_{0}$ many-body green’s function method, while the Bethe-Salpeter equation is solved to account for electron-hole interactions that play a crucial role in the optical properties of the family. A detailed symmetry analysis unravels the fine structure of excitons for all compounds. The exciton binding energy, excitonic wavefunctions and the dark-bright splitting are also reported for each material. It is shown that these quantities can be tuned over a wide range, form Wannier to Frenkel-type excitons, through for example substitutional engineering. In particular, Te-based materials, which share the electronic valency of corner-sharing Pb halide perovskites, are predicted to have exciton binding energies of above 1~eV and a dark-bright splitting of the excitons reaching over 100~meV. Our findings provide a fundamental understanding of the optical properties of the entire family of VODP materials and highlight how these are not in fact suitable Pb-free alternatives to traditional halide perovskites.
\end{abstract}

\vspace{0.6cm}

\clearpage
\newpage

Perovskites are some of the most common naturally occurring compounds in nature. These AMX$_3$ materials are made of three-dimensional networks of corner-sharing octahedra, and have been employed in a broad range of applications, like solar cells~\cite{Stranks2015}, highly-efficient light-emitters~\cite{Cui2021}, memories~\cite{Waser2009}, photo-catalysts~\cite{Zhang2019}, transistors~\cite{Xing2013} and superconductors~\cite{Reyren2007}. A$_2$MX$_6$ vacancy ordered double perovskites (VODP), are a particular sub-family of halide perovskites, which have recently emerged as promising materials for a broad range of applications \cite{Maughan2019,Ju2018}. In fact, the oxidation of Sn$^{+2}$ in its tetravalent state Sn$^{+4}$ leads to a structural transformation from the CsSnI$_{3}$ perovskite to the more stable VODP Cs$_{2}$SnI$_{6}$~\cite{Qiu2017}. To-date, the successful synthesis of many VODP materials have been reported, with tetravalent M-site atoms like Te~\cite{Maughan2016}, Ti~\cite{Chen2018}, Zr~\cite{Abfalterer2020}, Hf~\cite{Liu2022}, Pd~\cite{Sakai2017}, Pt~\cite{Evans2018}, Se~\cite{Wernicke1980}, Sb~\cite{Day1963}, Sn~\cite{Lee2014} and Ge~\cite{Laubengayer1940}. Most of these materials are stable and exhibit tunable electronic and optical properties making them targets of choice for optoelectronic applications, such as solar cells~\cite{Lee2014,Chen2018} and light-emitting devices~\cite{Abfalterer2020,Zhang2022,Holzapfel2020,Sun2022}. Among these properties, there is growing interest for the potential presence of strong excitonic effects in VODP \cite{Hoye2022,Rieger2019}, as the materials are made of isolated metal-halide octahedra, which are intrinsically structurally confined~\cite{Cucco2021}. Such excitonic properties are critical for the potential application of these photo-active materials, as for example they can limit charge carrier separation, or enhance photo-emission, hence open pathways for new opto-electronic devices that take advantage of such exciton physics. A deep understanding of the excitonic processes is necessary for probing the origins of the tunable and broad photoluminescence that is  observed in VODP~\cite{Chen2022,Liu2022}, which is commonly associated to excitons trapping. In fact, the formation of self-trapped excitons due to strong lattice distortions is one of the most discussed and debated underlying mechanisms~\cite{Zeng2021,Zheng2022,Zhang2022,Liu2022}. To-date, a detailed description of even the most fundamental excitonic features of the VODP  is still missing.

In this letter, we analyze the role of the electronic structure including electron-hole (e-h) interactions in the VODP materials by employing state-of-art ab-initio G$_{0}$W$_{0}$ and Bethe-Salpeter calculations. We unveil key details of their electronic band structure and show the impact of e-h interactions on their optical properties. Prototypical VODP materials, Cs$_{2}$SnX$_{6}$, Cs$_{2}$TeX$_{6}$ and Cs$_{2}$ZrX$_{6}$ (with X=Br, I), are categorised based on the electronic configuration of the tetravalent atom at the M-site with a valency comprised of an s, p and d closed-shell, respectively.~\cite{Maughan2016,Lee2014,Abfalterer2020}. The halogen atoms are identified as a tuning parameter to control the empty space in between the isolated metal-halide octahedra. We further show the effects of both M- and X-site substitutions on the electronic and optical properties by performing a symmetry analysis of the electronic structures. We find the most critical quantities of the photo-active VODP materials, such as the charge carrier effective masses, exciton binding energies and dark-bright exciton splitting and establish trends which allows the enhancement of these parameters via substitutional engineering. Te-based materials exhibit promising hole transport properties, while Sn-based VODP are most favoring electron transport. In addition, we show that Te-based materials on one hand exhibit promising electronic structure with an active lone pair, similar to Pb halide perovskites,  on the other exhibit drastically different excitonic features, which can be explained due to strong electron-hole interactions. Thus, we demonstrate that VODP even when sharing the same electron configuration with Pb halide perovskites, cannot be considered as their Pb-free alternatives. Finally, the absorption onset of all materials are shown to be dominated by electron-hole interactions leading to excitons ranging from Wannier-type to Frenkel-type. In all cases, the investigation of the exciton fine structure reveals that a non-optically active dark state lies below the first bright state. The electron-hole exchange interaction at the origin of the splitting between dark and bright exciton states is particularly strong in the case of Te-based materials. Interestingly, by choosing a metal and a halide, one can browse a range of exciton binding energies greater than 1 eV and of dark-bright splitting greater than 150 meV.

\begin{figure}[!ht]
 \begin{center}
  \includegraphics[width=1\textwidth]{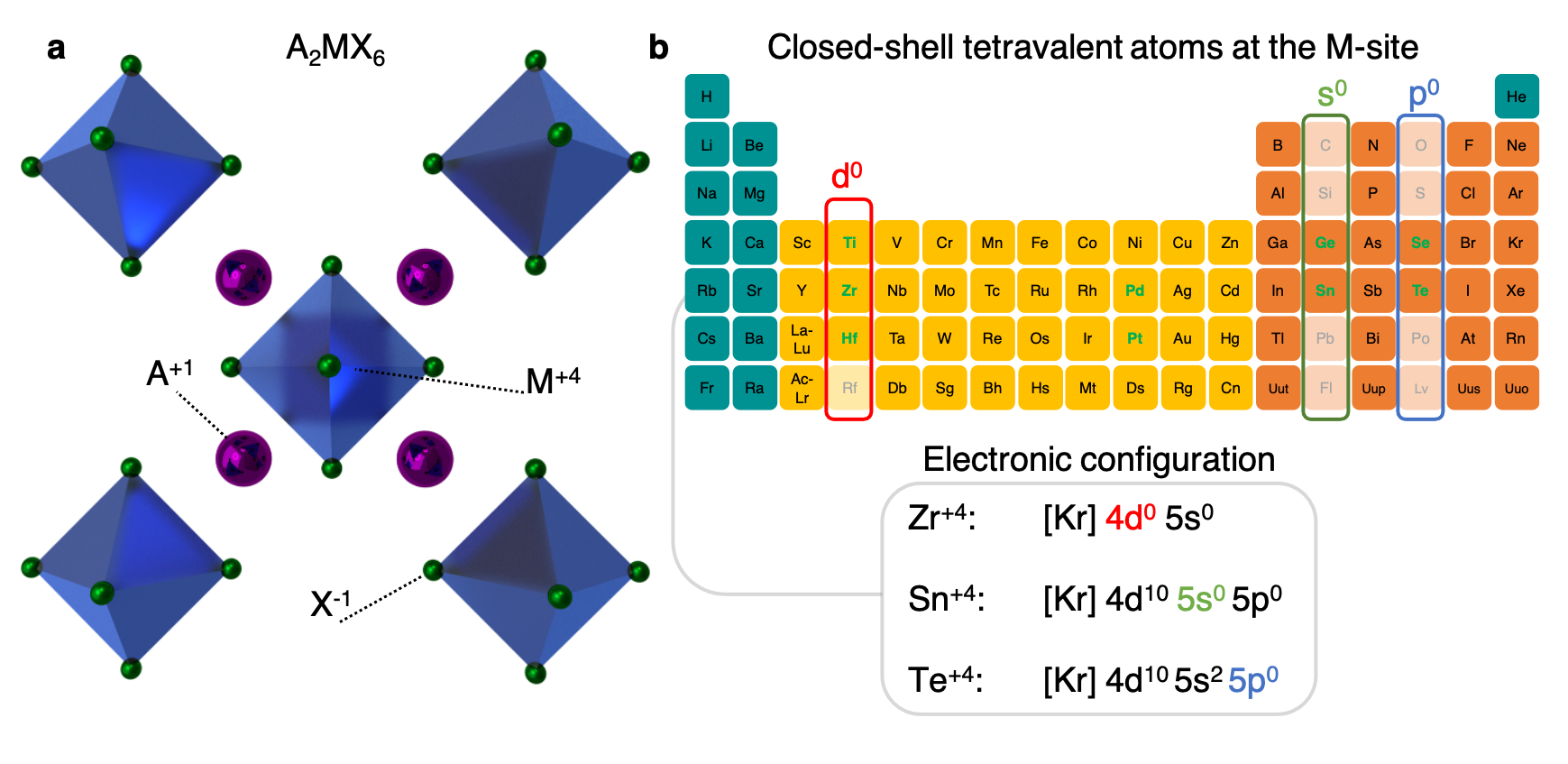}
 \end{center}
 \caption{\textbf{Vacancy ordered halide double perovskite family}. Representation of the VODP family of materials and chosen M$^{4+}$ metal sites. The boxes mark the s$^{0}$, p$^{0}$ and d$^{0}$ columns for a $+4$ oxidation state and electronic configuration of selected atoms. The atomic species in green have been synthesized as VODP with the marked atoms at the M$^{4+}$-site.}
 \label{fig:materials}
\end{figure}

VODP typically crystallize in the Fm$\bar{3}$m face-centered cubic lattice (space group number 225). The lattice corresponds to a rock salt arrangement of corner-sharing MX$_{6}$ and $\Delta$X$_{6}$ octahedra with $\Delta$ being the vacancy site, or simply a double perovskite for which one of the two M-sites is vacant, as show on Figure~\ref{fig:materials}a.
We sample the VODP family by selecting prototypes based on the nominal valency (i.e. electronic configuration) of the tetravalent cation at the M-site. In Figure~\ref{fig:materials}b we highlight the columns of the periodic table that correspond to atoms with s-p-d filled closed-shells at a +4 oxidation state. The atomic species that are highlighted in each closed-shell column were successfully employed to form stable VODP. Here we select period 5 of the periodic table, thus Sn, Te and Zr as the atomic species at the M-site of each closed-shell VODP type, generating 6 different materials Cs$_{2}$TeX$_{6}$, Cs$_{2}$SnX$_{6}$, and Cs$_{2}$ZrX$_{6}$ (with X=Br, I). Interestingly, Te$^{4+}$ is isoelectronic with the Pb$^{+2}$ ion, which is the building block of the most efficient perovskite materials for optoelectronic applications. As a starting point, the atomic coordinates and lattice parameters for these materials are fully optimized using ab-initio calculations based on Density Functional Theory (DFT) as detailed in the supplementary information file (SI). Table~S1 of the SI summarizes the obtained lattice parameters, metal-halogen bond lengths and volumetric size of the vacancy. The optimized structures for all compounds are in the Fm$\bar{3}$m space group, except Cs$_{2}$SnI$_{6}$, which exhibits very slightly rotated Sn-I octahedra with small octahedra distortions, in agreement with the work of Jong \textit{et al.}~\cite{Jong2019}. As previously shown in the context of d$^{0}$ VODP, halogen-halogen \textit{p}-orbital interactions dictate the empty space formed by the vacancies in these materials~\cite{Cucco2021}. Here the same is observed for all VODP regardless of the atom at the M-site, and the volume of the empty space in between the MX$_6$ octahedra depends solely on the halogen size. Hence, Br to I substitution can be used as a tuning parameter to control the size of the vacancy from 33 to 40~\AA$^3$, as defined in Table~S1. As a final assessment of the mechanical stability of the optimized structures, we calculate the phonon dispersion for each material. These are shown in Figure~S1, where the absence of phonon imaginary modes in all studied compounds indicates mechanical stability. The redshift of the frequency that is observed when going from bromides to iodides is consistent with the increase of halogen's atomic mass and thus damping of the M-X oscillation modes.

\begin{figure}[!ht]
 \begin{center}
  \includegraphics[width=1\textwidth]{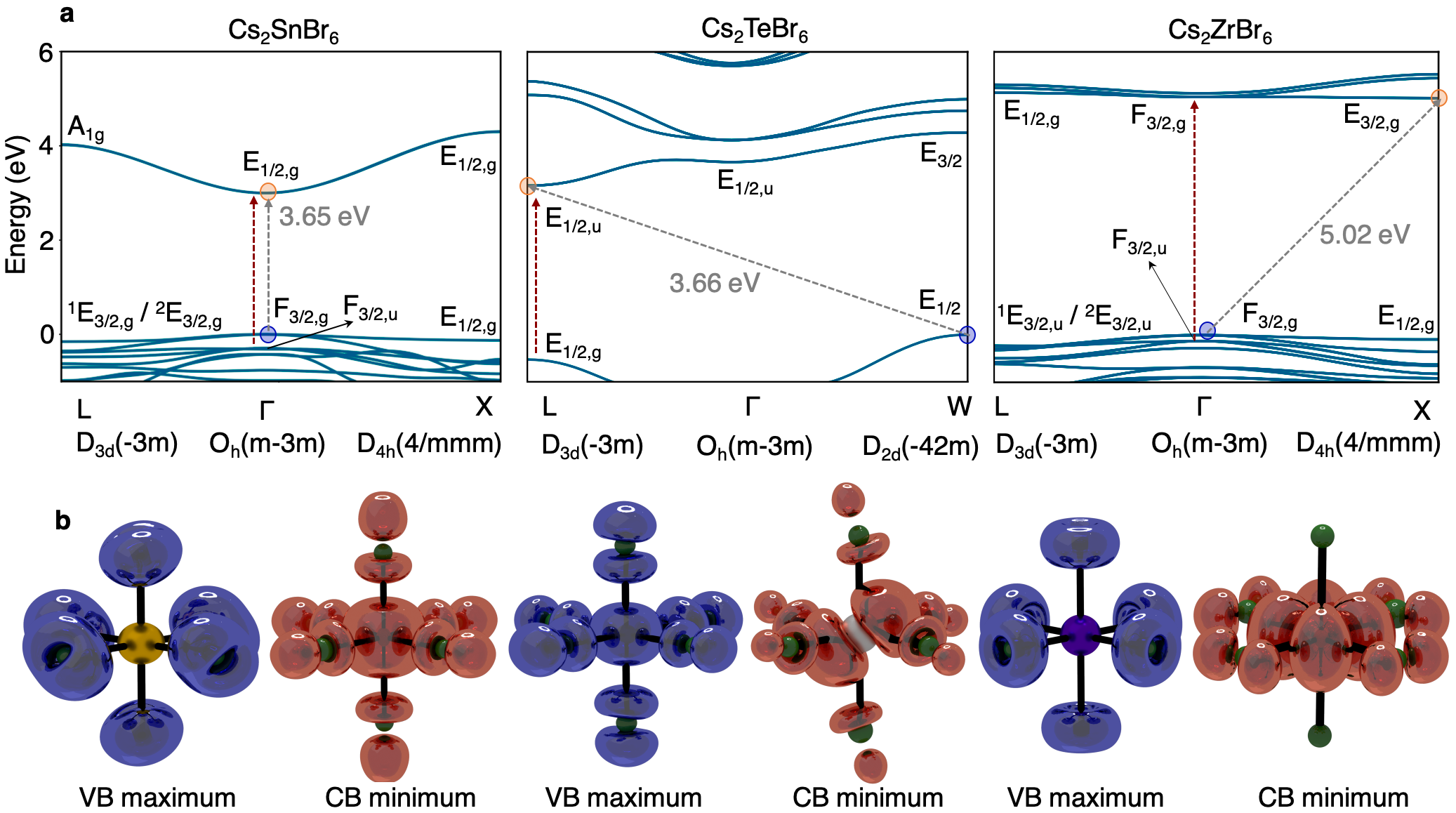}
 \end{center}
 \caption{\textbf{G$_{0}$W$_{0}$ electronic band structures and charge density}. (a) G$_{0}$W$_{0}$ electronic band structures and charge densities. The irreducible representations and point group symmetries are shown and the valence band maximum (VBM) and conduction band minimum (CBM) are marked with blue and orange circles, respectively. The grey and red arrows represents the fundamental band-gap and first direct allowed transition, respectively. (b) Charge densities calculated at the VBM and CBM for Cs$_{2}$MBr$_{6}$ (M=Sn, Te, Zr). }
 \label{fig:bands}
\end{figure}

Two critical parameters for the accurate calculation of excitonic properties are the band-gaps and charge-carrier effective masses, as both can have strong impacts on the exciton binding energies and the relative position of excitonic peaks. To this aim, we employ G$_{0}$W$_{0}$ calculations that take into account quasi-particle effects due to screened electron-electron interactions. Figure~\ref{fig:bands}a shows the electronic band structures for the bromides, while the electronic bands of the iodides can be found in Figure~S2a of the SI.
Sn-based compounds exhibit direct band-gaps at the $\Gamma$ point. The optical transition from the valence band maximum (VBM) to conduction band minimum (CBM) is given by a transition from the irreducible representation F$_{3/2,g}$ to E$_{1/2,g}$ which is parity forbidden. The first direct dipole allowed transition for the Sn-based compounds is from VBM-1 (F$_{3/2,u}$) to CBM (E$_{1/2,g}$), i.e. 300~meV and 350~meV larger than the fundamental band-gap of Cs$_{2}$SnBr$_{6}$ and Cs$_{2}$SnI$_{6}$, respectively. Due to the different orbitals involved at the band edges, Te-based compounds have indirect band-gaps between the L-W high symmetry points. Interestingly, the shapes of the Cs$_{2}$TeX$_{6}$ valence band and the Cs$_{2}$SnX$_{6}$ conduction band are the same, which is consistent with the electronic configurations [Kr]5s$^{\textrm{2}}$ and [Kr]5s$^{\textrm{0}}$ for Te$^{4+}$ and Sn$^{4+}$, respectively. We note that within DFT-PBE the indirect band-gap of Cs$_{2}$TeI$_{6}$ is incorrectly predicted to be located between $\Gamma$-L (see Figure~S3). Within G$_{0}$W$_{0}$ Te-based compounds exhibit an indirect band-gap between L-W, which we also verified by employing PBE0 hybrid functional. Based on symmetry analysis we identify the first direct dipole allowed transition as being from the VBM (E$_{1/2,g}$) to the CBM (E$_{1/2,u}$) at L for Cs$_{2}$TeX$_{6}$. Within G$_{0}$W$_{0}$ these transitions are of 2.83~eV and 3.90~eV for Cs$_{2}$TeI$_{6}$ and Cs$_{2}$TeBr$_{6}$, respectively. For the Zr-based we identify the first direct dipole allowed transition as being from the VBM-1 (F$_{3/2,u}$) to CBM (F$_{3/2,g}$). Within G$_{0}$W$_{0}$ these transitions are of 5.18~eV and 3.64~eV for Cs$_{2}$ZrBr$_{6}$ and Cs$_{2}$ZrI$_{6}$, respectively.

The electronic charge density at the VBM and CBM, are shown in Figure~\ref{fig:bands}b and Figure S2b for bromides and iodides, respectively. The VBM for the Sn-based and Zr-based materials is made of halogen $p$-orbitals, and there is no contribution of the M-site atoms. This is not the case for Te-based compounds where the halogen $p$-orbitals interact with the Te $s$-orbitals forming a VBM that resembles in character traditional Pb-based halide perovskites~\cite{Volonakis2017}. This is of potential interest for the hole transport within this materials, as only the Te-based compounds have a uniform density that expands in three-dimensions at the VBM. The CBM for Sn-based materials is made of Sn $s$-orbitals hybridized with halogen $p$-orbitals, hence is similar to the VBM of Te-based materials. The CBM of Te-based materials is made of Te $p$-orbitals and halogen $p$-orbitals, while for the Zr-based materials we find Zr $d$-orbitals and halogen $p$-orbitals. In terms of electron transport Sn-based and Te-based materials are the most promising as the $s$- and $p$-orbitals of the M-site atoms are involved in the CBM. Overall, the contribution of the atoms at the M-sites to the band edges is in perfect agreement with their electronic valency at their nominal oxidation state, which was described in the s$^0$, p$^0$ and d$^0$ families of Figure~\ref{fig:materials}b. Table~S2 contains the electronic band-gaps within DFT and G$_{0}$W$_{0}$.
\begin{table}[H]
\centering
\caption{Average electron (m$_e$) and hole (m$_h$) effective masses at the band edges within G$_{0}$W$_{0}$. Numbers between parenthesis represents the heavy bands effective masses. The complete effective mass components can be found at Table~S3 of SI.}
\label{tab:2}
\begin{tabular}{@{}ccccccc@{}}
\toprule
\begin{tabular}[c]{@{}l@{}}Effective\\ masses(m$_{0}$)\end{tabular}  & Cs$_{2}$SnBr$_{6}$ & Cs$_{2}$SnI$_{6}$ & Cs$_{2}$TeBr$_{6}$ & Cs$_{2}$TeI$_{6}$ & Cs$_{2}$ZrBr$_{6}$ & Cs$_{2}$ZrI$_{6}$ \\ \midrule
m$_{e}$  & 0.24 & 0.29 & 0.33 & 0.39 & 1.80 & 1.24     \\
m$_{h}$  & 0.55 (1.22) & 0.58 (1.08) & 0.46 & 0.29 & 1.12 (2.82) & 0.60 (1.45)    \\ \bottomrule
\end{tabular}
\end{table}

To quantify the band dispersion we calculate the charge carrier effective masses at the band edges within G$_{0}$W$_{0}$, as summarized in Table~\ref{tab:2}. The Sn materials exhibit the lightest and isotropic electron effective masses, which are consistent with the M-site s-orbital character involved in their electronic structure. Te compounds exhibit particularly light masses for both holes and electrons, as these involve M-site s-orbitals and p-orbitals at the valence and conduction band, respectively. These values confirm the potential good electron carrier transport properties of the Sn-based and Te-based VODP, while for hole transport only Te-based materials would not be limited by heavy masses at the VBM. In fact, both Sn-based and Zr-based materials exhibit light and heavy holes at the VBM, which would challenge their potential performance.

\begin{figure}[!ht]
 \begin{center}
  \includegraphics[width=\textwidth]{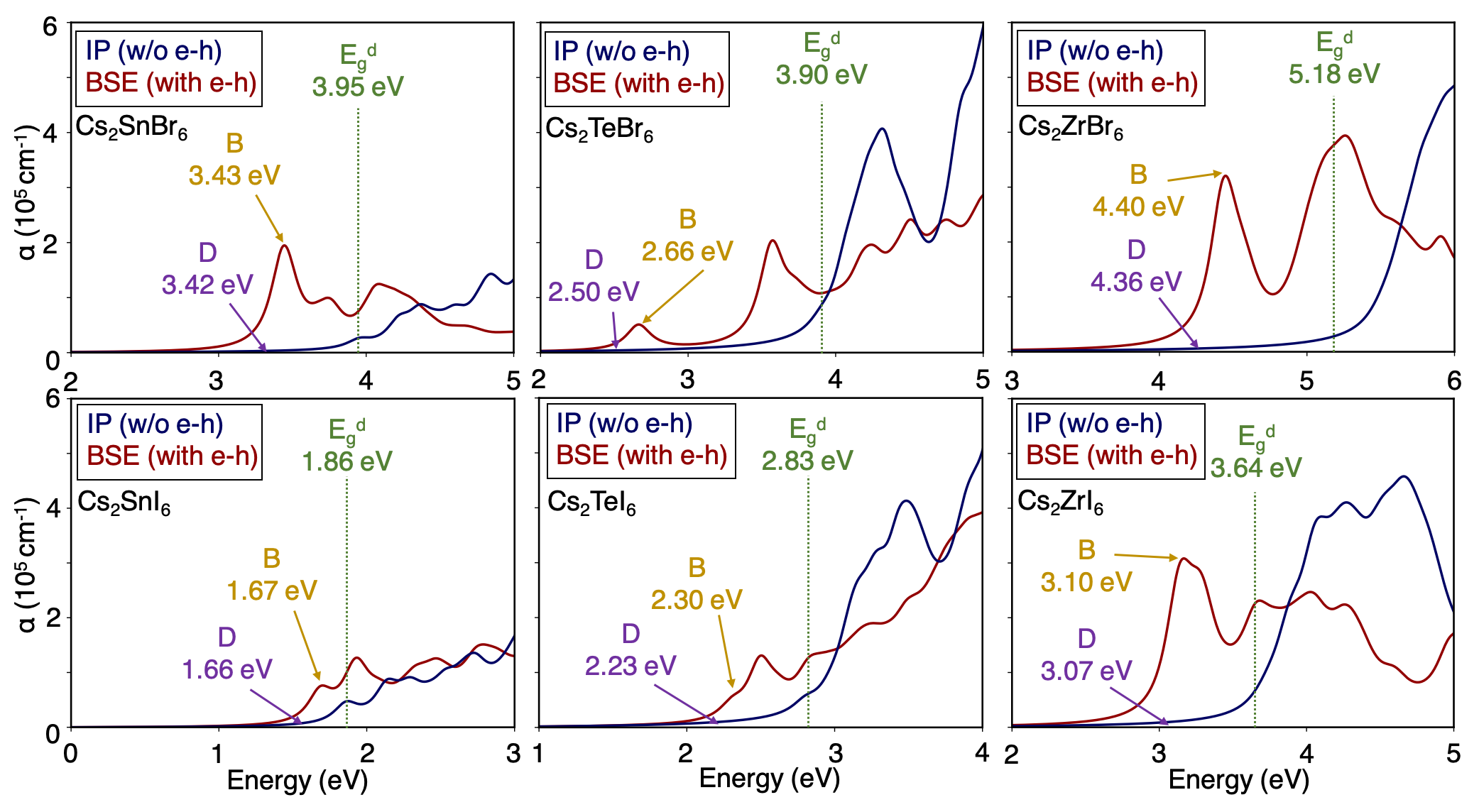}
 \end{center}
 \caption{\textbf{Optical absorption}. Optical absorption coefficient of the Cs$_{2}$MX$_{6}$ (M=Sn, Te, Zr and X=Br, I) vacancy ordered perovskite family within BSE@G$_{0}$W$_{0}$ and IP@G$_{0}$W$_{0}$ level. The red and blue curves show the BSE (with electron-hole interactions) and IP (without electron-hole interactions) coefficients, respectively. The positions of the dark (D) and bright (B) exciton states are shown. The green dotted lines show the position of the first direct dipole-allowed transition E$_{g}^{d}$ at G$_{0}$W$_{0}$ level.}
 \label{fig:optics}
\end{figure}

\begin{figure}[!ht]
 \begin{center}
 \includegraphics[width=0.85\textwidth]{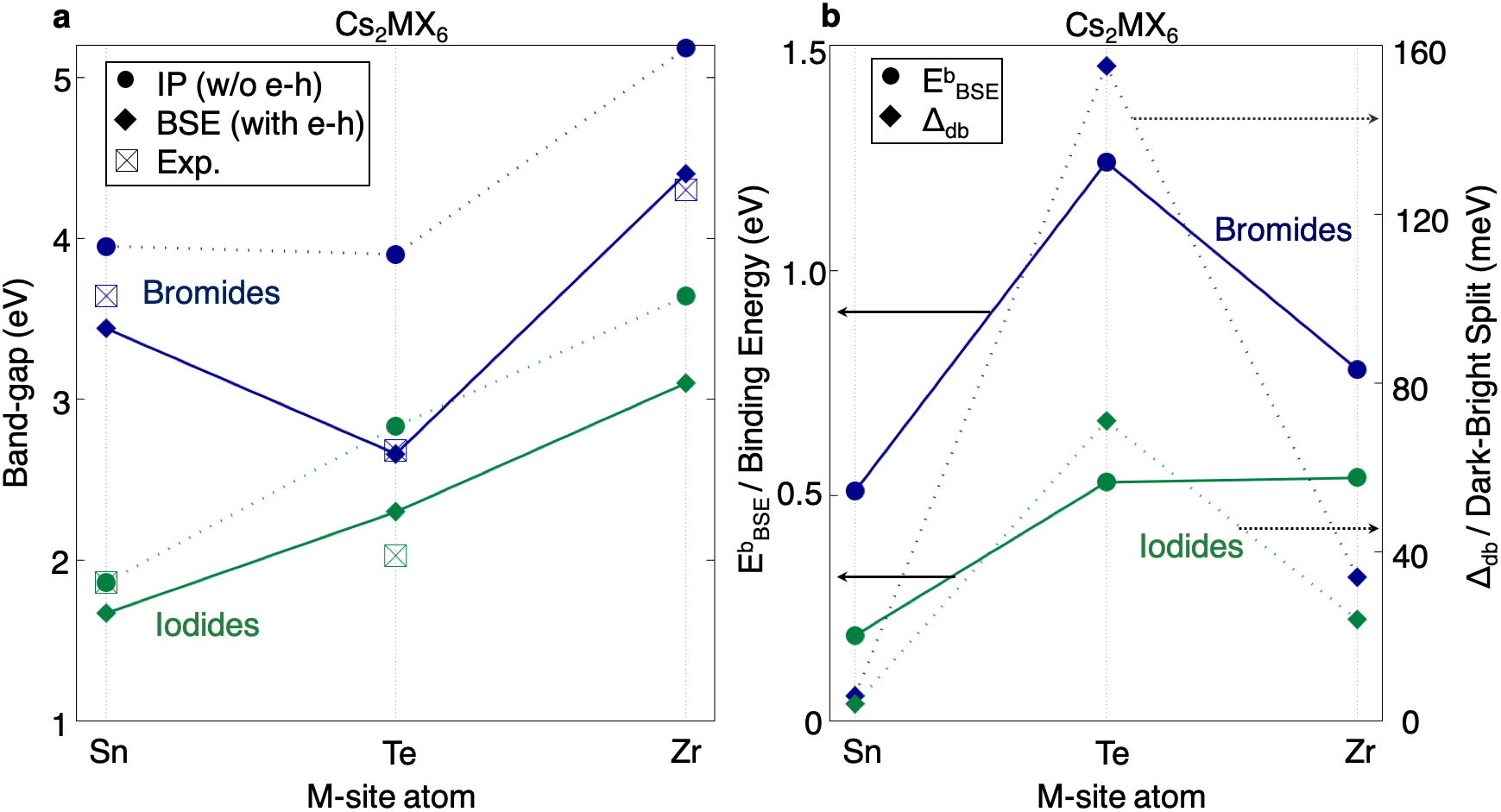}
 \caption{\textbf{Band-gaps, exciton binding energy, and dark-bright splitting:} (a) Comparison between the first dipole-allowed transition at G$_{0}$W$_{0}$ level, first bright peak of BSE calculation, and experimentally measured optical band-gaps. (b) Binding energies E$^{b}$ and dark-bright splitting $\Delta_{db}$ trend obtained from the BSE calculations. Blue and green curves represent bromides and iodides, respectively.}
 \label{fig:plots}
 \end{center}
\end{figure}

We move to evaluate the materials optical properties by employing the ab-initio Bethe-Salpeter formalism~\cite{Salpeter1951} starting from our G$_{0}$W$_{0}$ calculations (BSE@G$_{0}$W$_{0}$), which allow taking into account the band-gap renormalization and dispersion corrections over standard DFT. The BSE@G$_{0}$W$_{0}$ approach allows including e-h coupling interactions, thus the investigation of bounded excitonic states.
%
%
Figure~\ref{fig:optics} shows the absorption coefficients obtained (i) at the BSE@G$_{0}$W$_{0}$, and (ii) without e-h interactions at the independent particle (IP) at the G$_{0}$W$_{0}$ level. Due to the inclusion of e-h coupling, all the BSE spectra red-shift and the oscillator strengths are redistributed when compared to the IP calculations. For all materials we find a bright exciton state well within the band-gap, and dark states (not optically active) at lower energies. Furthermore, there is clearly a very strong modification to the spectra when e-h interaction are included, which is consistent with the isolated octahedra that make the VODP lattice. To further analyse the optical spectra, we first look at the optical band-gaps. Figure~\ref{fig:plots}a, shows the first dipole-allowed transitions (optical band-gap) with and without e-h interactions, and the experimental data. BSE values are in excellent agreement with available measurements, which we extract by picking the first excitonic peak in the absorption of references~\citenum{Kavanagh2022,Folgueras2021,Vazquez2020,Abfalterer2020}.

Next, we define the exciton binding energy ($E^{b}_{BSE}$) as the energy difference between the first allowed transition and the bright exciton peak. As shown in Figure~\ref{fig:plots}b, we report large binding energies ranging from 190~meV to 540~meV for iodine-based and 510~meV to 1240~meV for bromine-based compounds. These values are particularly large as can be expected from the quasi-0D nature of the weekly interacting isolated octahedra~\cite{Cucco2021,Maughan2019}.
Furthermore, the exciton binding energy is systematically larger for bromide compounds than their iodide counterparts. This effect is related to the connection between the electronic band-gap and the dielectric constant of these materials, as the electron polarizability decreases as the band-gap increases. Hence, the screening of the Coulomb interactions between electrons and holes becomes less efficient leading to tightly bounded excitonic states~\cite{Marongiu2019}. This is consistent with the literature for excitons in AMX$_{3}$ halide perovskites~\cite{Galkowski2016,Marongiu2019,Jiang2017} and also A$_{2}$MM'X$_{6}$ halide double perovskites~\cite{Palumno2020,Biega2021}. A surprising finding is the very large $E^{b}_{BSE}$ for the Te-based materials, which share common optical features to Pb halide perovskites, such as strong optical transitions that involve $s$ (i.e. active lone-pair) to $p$ M-site orbitals. Yet, Pb halide perovskites exhibit low exciton binding energy.\cite{Rana2019,Miyata2015,Marongiu2019} Such strong e-h interactions observed in the case of the Te-based materials, can be attributed to their intrinsic structural confinement and their indirect electronic structure, that give rise to  Frenkel-like excitons. Hence, despite their similar valency Te-based VODP are far from being considered as alternatives to Pb-based halide perovskites.

\begin{figure}[!ht]
 \begin{center}
  \includegraphics[width=\textwidth]{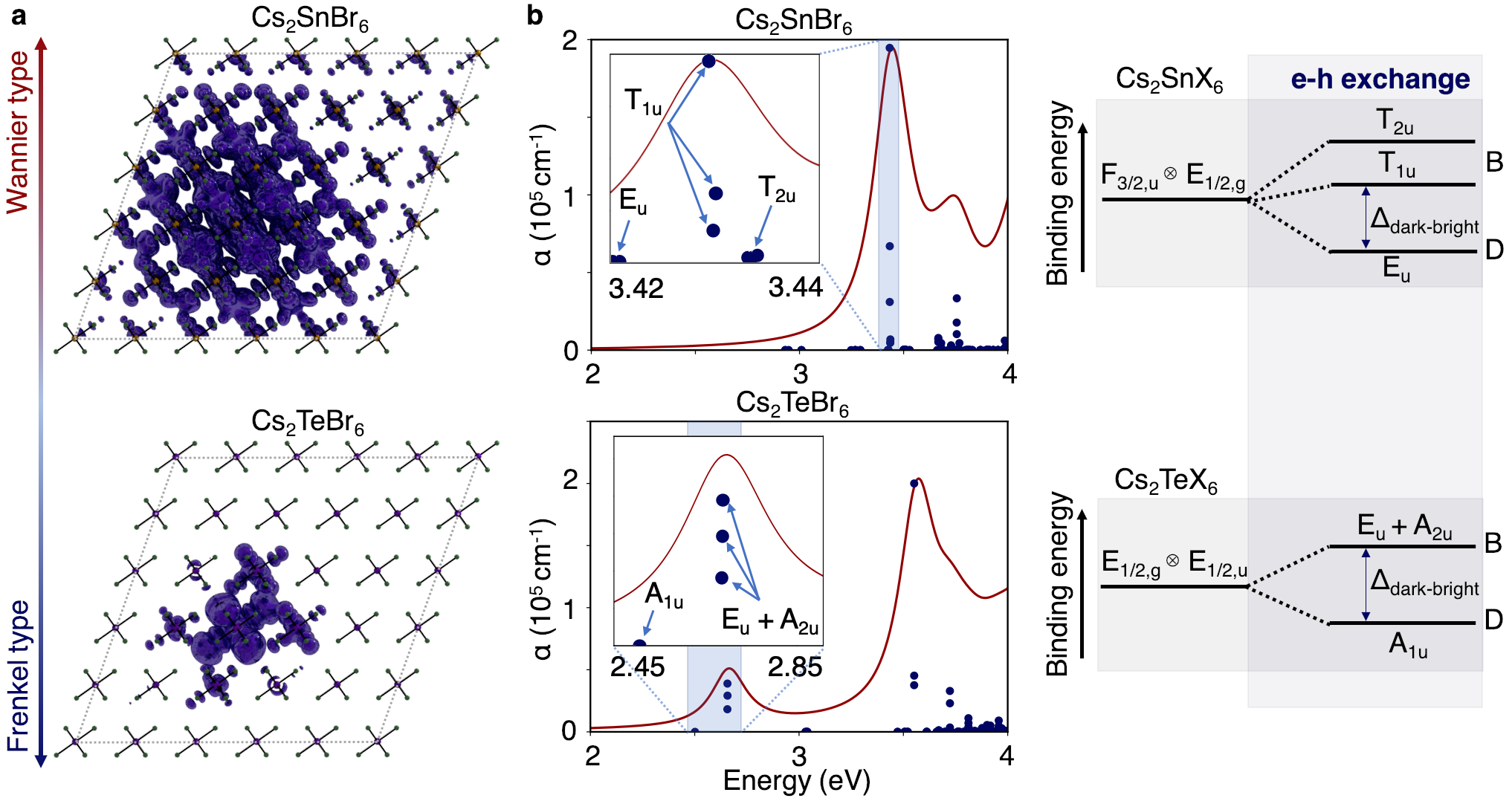}
 \end{center}
 \caption{\textbf{Exciton spatial distribution and fine structure}. (a) Exciton spatial distribution $|\Psi(r_{h},r_{e})|^{2}$ for the bright exciton. The hole position r$_{h}$ is fixed in the vicinity of the halogen atoms. The isovalue is set to represent 85\% of the total distribution. (b) Calculated exciton oscillator strength with fine structure shown as inset plot. On the right, an illustration of the dark-bright splitting.}
 \label{fig:trends}
\end{figure}

The exciton wavefunction for the first bright exciton is visualised in Figure~\ref{fig:trends}a, as a spatial distribution given by $|\Psi(r_{h},r_{e})|^{2}$. The exciton wavefunction is more localized in the case of Te-based VODP, while Cs$_{2}$SnBr$_{6}$ exhibits a significantly delocalized excitonic wave-function which extends over several unit-cells of the lattice. This effect is even stronger for the case of iodides shown in Figure~S4,  in agreement with the calculated exciton binding energies. In fact, the exciton in Cs$_{2}$SnI$_{6}$ is closer to the Wannier-type, which is also confirmed by its E$_{BSE}^{b}$ of 190~meV sizeably smaller than for the other compounds. On the other hand, Cs$_{2}$TeX$_{6}$ and Cs$_{2}$ZrX$_{6}$ compounds are closer to a Frenkel-type description, exhibiting stronger exciton localization. Over the series of closed-shell VODP, there is an interplay between Wannier and Frenkel-type excitons that is related to the atomic orbital at the M-site, as a clear Wannier-like exciton is formed for Cs$_{2}$SnI$_{6}$ and a Frenkel-type for Cs$_{2}$TeBr$_{6}$. This trend can be followed across the series as shown with the projection of the exciton weights $|A^{\lambda}_{eh}|$ over the band-structure, where $|\lambda\rangle = \sum_{eh}A^{\lambda}_{eh}|eh\rangle$ shown in Figure~S5. The exciton weights in Sn-based materials are close to the band edges (delocalised excitons), less so for the case of Zr and Te iodides (localised), while for Zr and Te bromides these expand over the entire Brillouin zone (strongly localised). The materials order from Wannier-like to Frenkel-type excitons perfectly follows the $E^{b}_{BSE}$ trend. In addition, we compare our ab-initio results with a semi-empirical Wannier-Mott model (see SI for details), and we show that the model fails to qualitatively assess the relationship between the exciton binding energies for all cases but the Sn iodide material.

Finally, we investigate the detailed symmetry and fine structure of the excitons. As shown in Figure~S5, the bright exciton of Cs$_{2}$SnBr$_{6}$ is formed mainly by a localized transition around $\Gamma$ from the VBM-1 to CBM, and a small contribution from the optically forbidden transition from VBM to CBM, while for Cs$_{2}$SnI$_{6}$ this contribution from the VBM is not observed. The symmetry of the exciton is given by a direct product between the irreducible representations of the corresponding electron-hole pairs, i.e. $F_{3/2,u}\otimes E_{1/2,g}=E_{u}\oplus T_{1u} \oplus T_{2u}$ multiplied by a s-like envelope function (i.e. for s-like excitons, and hence doesn't change the symmetry). The bright exciton of Cs$_{2}$TeX$_{6}$ are mainly composed by transitions VBM to CBM at L but with smaller contributions around the W high-symmetry point. Symmetry analysis at the L point gives $E_{1/2,g}\otimes E_{1/2,u}=E_{u}\oplus A_{2u} \oplus A_{1u}$. Finally, for Cs$_{2}$ZrX$_{6}$ the exciton is formed by transitions from VBM to CBM at the $\Gamma$ point, giving a symmetry of $F_{3/2,u}\otimes F_{3/2,g}=A_{1u}\oplus E_{u} \oplus 2T_{1u}\oplus 2T_{2u}\oplus A_{2u}$. We also observe contributions from the transition VBM to CBM+1 for the bromide compound, due to the CBM and CBM+1 be closer in energy when compared to iodides.

The bright exciton states are in fact formed by electrons and holes with anti-parallel spins and can recombine easily thought the emission of a photon, and thus exhibit fast lifetimes~\cite{Zhang2017}. Yet, a bounded excitonic state with parallel spins can also exist and due to spin momentum conservation these cannot recombine via the direct emission of a photon and therefore exhibit a larger radiative lifetime when compared to bright excitons. These so called dark excitons are not intrinsically optically active, i.e. exhibit zero oscillator strength, and arise from the electron-hole exchange interaction. These states and their energy split from the bright states ($\Delta_{db}$), are of particular interest for state-of-art technologies such as optically controlled information processing and fast/slow light applications.~\cite{Zhang2017} Figure~\ref{fig:plots}b and Table~S4 summarizes the obtained $\Delta_{db}$ values for all compounds within this work, while Figure~\ref{fig:trends}b presents the detailed exciton fine structure for the Sn- and Te-based compounds. From the symmetry and oscillator strength analysis of Cs$_{2}$SnX$_{6}$, we find an $E_{u}$ doublet dark exciton state, which is split from the bright triplet peak $T_{1u}$ by 6~meV and 4~meV for bromide and iodide, respectively. The splitting between the bright peak $T_{1u}$ and the optically inactive triplet state $T_{2u}$ is also observed. For the Te-based compounds, we find an $A_{1u}$ dark singlet exciton which is split from the bright peak comprised of the optically active doublet $E_{u}$ and singlet $A_{2u}$. The dark-bright splitting are of 155~meV and 71~meV for bromide and iodide Te-based compounds. The energy difference between the $E_{u}$ and $A_{2u}$ is below the resolution of our calculation and as so, are treated as practically a triply degenerated state. For the fine structure of Zr-based compounds, due to the large number of energetically close electronic states near the VBM a strong coupling of the valence bands and also a large number of transition states are expected. This leads to several oscillator strengths in the vicinity of the bright excitonic peak, which do not allow the proper identification of the sixteen $A_{1u}\oplus E_{u} \oplus 2T_{1u}\oplus 2T_{2u}\oplus A_{2u}$ states, as shown in Figure~S6. Yet, by analysis of the e-h pairs contributions of Figure~S5 we identify the dark state as being originated from a VBM to CBM transition at the $\Gamma$ point. The dark-bright splitting for these are 34~meV and 24~meV for bromide and iodide, respectively. $\Delta_{db}$ decreases from Te, Zr to Sn, with Te-based compounds exhibiting the larger dark-bright splitting when compounds with the same halogen are compared. Moreover, bromine-based compounds are shown to exhibit a larger splitting than iodine-based ones, especially in the case of Cs$_2$TeX$_6$. To further explore this, we solve the BSE without including the e-h exchange interaction $\bar{V}^{vck}_{v'c'k'}$ (see SI). Figure~S7 shows that the spectrum renormalization caused by the e-h exchange interaction is always greater for bromides than for iodides. Interestingly, this effect has also been observed for CsPbX$_{3}$ halide perovskites nanocrystals~\cite{Becker2018}, for which the larger exciton binding energy of bromides is attributed to the larger e-h exchange interactions. The effect of the interaction appears particularly strong in the case of Te compounds. In the case of VODP, both metal and halogen substitution allow tuning $\Delta_{db}$
within a considerable range from 4~meV to 155~meV.

In conclusion, we investigated the structural and optical properties of the VODP family of materials. We show that the volume of the vacant site can be tuned solely by the halogen composition. Based on their electronic structure we find Te-based compounds to be most interesting for hole transport, while Sn-based compounds are suitable for electron transport. As expected for structurally confined materials, it is shown that electron-hole interactions play a major role on the optical properties of these compounds. Indeed, they exhibit large and tunable exciton binding energies through a range of 0.2-1.2~eV by both M- and X-site composition, with bromides exhibiting larger binding energies than iodides. By employing symmetry analysis of the electronic band structures, we elucidate the bright-exciton symmetries and fine-structure as to obtain the dark-bright exchange splitting and show its tunability across the family. The electron-hole exchange interaction is shown to have major contribution on the dark-bright splitting, in particular for Te-based compounds. Finally, by a joint analysis of the exciton wavefunctions and binding energies we conclude that the VODP family can host both Wannier- and Frenkel-type excitons depending on the choice of M- and X-site species. Our findings provide fundamental understanding of the optical properties of the prominent VODP family of materials.
It exhibits intrinsically different excitonic properties with respect to APbX$_3$ materials, even when sharing electronic configuration. Consequently, due to the large and/or indirect electronic band gaps, heavy effective masses of the charge carriers, and the large exciton binding energies these materials cannot be the base of alternative Pb-free photovoltaic devices, though they present great opportunity for white-light emission applications and quantum optical technologies.

\section*{Acknowledgments}
We gratefully acknowledge discussions with Prof. Feliciano Giustino, Dr. Joshua Leveillee and Dr. Tan Nguyen. The research leading to these results has received funding from the Chaire de Recherche Rennes Metropole and the European Union’s Horizon 2020 program, through a FET Open research and innovation action under the grant agreement No 862656 (DROP-IT). This work was granted access to the HPC resources of TGCC under the allocations 2020-A0100911434 and 2021-A0110907682 made by GENCI. We acknowledge PRACE for awarding us access to the ARCHER2, United Kingdom.

\section*{Support Information File}
Computational details; Structural parameters; Summary of Band-gaps; Charge carrier effective masses; Parameters of Wannier-Mott model; Binding energies and dark-bright splitting; Phonon dispersion; Electronic structure of iodides; Projected density of states; BSE calculations with and without exchange interaction; Exciton spatial distribution for iodides and Cs$_{2}$ZrBr$_{6}$; Projection of excitonic weights over band-structure; Exciton fine structure of Cs$_{2}$ZrBr$_{6}$; CIF files for all  six materials.

\clearpage

{\bf References\\}
\bibliographystyle{naturemag}
\bibliography{bibliography}

\end{document}